\documentstyle[aps,prbbib,twocolumn,epsf]{revtex}

\begin{document}
\draft
\title{Nonlinear transport theory for hybrid normal-superconducting devices}

\author{Jian Wang$^1$, Yadong Wei$^1$, Hong Guo$^2$, Qing-feng Sun$^3$, and 
Tsung-han Lin$^3$} 
\address{1. Department of Physics, The University of Hong Kong, 
Pokfulam Road, Hong Kong, China\\
2. Center for the Physics of Materials and Department 
of Physics, McGill University, Montreal, PQ, Canada H3A 2T8\\
3. State Key Laboratory for Mesoscopic Physics and Department 
of Physics, Peking University, Beijing 100871, China}
\maketitle

\begin{abstract}
We report a theory for analyzing nonlinear DC transport properties 
of mesoscopic or nanoscopic normal-superconducting (N-S) systems. 
Special attention was 
paid such that our theory satisfies gauge invariance. At the linear 
transport regime and the sub-gap region where the familiar scattering 
matrix theory has been developed, we provide confirmation that our theory 
and the scattering matrix theory are equivalent. At the nonlinear regime, 
however, our theory allows the investigation of a number of important 
problems: for N-S hybrid systems we have derived the general nonlinear 
current-voltage characteristics in terms of the scattering Green's 
function, the second order nonlinear conductance at the weakly nonlinear 
regime, and nonequilibrium charge pile-up in the device which defines the 
electrochemical capacitance coefficients. 

\end{abstract}


\section{Introduction}

The physics associated with quantum conduction in various low dimensional 
hybrid superconducting systems has been a major focus of research at 
present\cite{paper1,beenakker,been1,lambert}. Due to advances in controlled 
crystal growth and lithographic techniques, it is now possible to fabricate 
various submicron hybrid superconducting structures where accurate 
experimental measurements can be made\cite{experiment}. An important 
theoretical task is to be able to predict, for general mesoscopic or 
nanoscopic hybrid systems, transport 
properties such as the nonlinear current-voltage characteristics and the 
nonequilibrium charge distribution inside the system as a function of the 
applied bias voltage. Our theoretical understanding of quantum transport 
in these very small N-S hybrid systems has been achieved by scattering 
matrix theory\cite{beenakker,lambert} and by non-equilibrium Green's 
function (NEGF) theory\cite{yeyati1,yeyati2,datta2,sun1,sun2}.

To analyze nonlinear transport coefficients, {\it i.e.} coefficients which
appear in front of nonlinear powers of bias voltage, in principle one must
make sure gauge invariance of the theory. This means that theoretical 
results should not change when bias voltage applied at all the device leads 
is changed by the same amount. This is a necessary condition for any 
transport theory and has been recognized in the literature\cite{but1}. 
Consider a device which is connected to the outside world by several leads 
$\alpha$ where bias voltage $V_\alpha$ is applied. When 
$V_\alpha \rightarrow V_\alpha+v$ where $v$ is a constant,
the calculated results (such as current) will not change if the
electrostatic potential $U$ inside the device is also changed by the same
amount $v$. However $U=U({\bf r})$ which is the Hartree potential, can
only be obtained by solving a self-consistent problem. In other words,
to satisfy gauge invariance one is necessarily required to consider 
Coulomb interactions at least at the Hartree level. Furthermore, 
in general when external bias voltage is applied to a device, the 
flow of charge carriers through the device could polarize the system due to 
long range Coulomb interactions. For a macroscopic metallic conductor, the 
polarization can be safely neglected since interaction is well screened.
However for mesoscopic scale and nano-scale conductors the polarization
could be very important. This also requires self-consistent analysis.

For {\it normal conductors}, B\"uttiker and his co-workers\cite{but1,but0} 
developed an approach based on the scattering matrix theory to deal with 
the second order nonlinear conductance coefficients. This theory can also 
be extended to higher nonlinear order in DC situations\cite{ma}. On the
other hand for a N-S hybrid system, despite the many theoretical 
investigations on its quantum transport property\cite{paper2}, the 
important issue of gauge invariance has not been clearly addressed so far. 
In light of this unsatisfactory situation, in this paper we report the 
development of a proper nonlinear transport theory which satisfies gauge 
invariance for mesoscopic or nanoscopic N-S hybrid device systems. Our 
theoretical 
formulation is based on nonequilibrium Green's function approach where the 
quantum transport problem is solved in a self-consistent manner. We have 
derived analytical expressions for the general I-V characteristics 
including the sub-gap behavior of the N-S device in terms of Green's 
functions which are numerically calculable. In the weakly nonlinear 
regime we further derived the second order nonlinear coefficient by 
solving the characteristic potential. Since our theory can deal with 
charge polarization in the presence of transport\cite{but4,wang2}, we 
have also derived the linear and second order nonlinear charge 
distributions which define the electrochemical capacitance of the system. 
Finally, at linear order for which scattering matrix theory (SMT) has been 
developed for N-S systems, we prove by explicit calculation that our 
theory and the SMT theory are equivalent. This way the widely used SMT 
theory and the NEGF theory are formally connected for N-S systems through 
a Fisher-Lee relation which we derive in this work. 

The rest of the paper is organized as follows. Section II and III present
the gauge invariant nonlinear transport theory for the N-S system and 
confirming its equivalence to the familiar scattering matrix theory at 
the linear transport regime.  Section IV provides several applications of
our theory.  Section IV is a short summary for this work.

\section{Gauge invariant theory}

In this section we develop the gauge invariant nonlinear transport theory
based on NEGF for N-S hybrid device systems. To be specific, the N-S system 
we consider is a quantum well connected to a normal metal lead and a 
superconducting lead.  It is described by the following 
Hamiltonian\cite{yeyati1,ando,sun1} in the second quantized form,
\begin{equation}
H=H_L + H_R + H_{d} +H_T ~~\ \ \ ,
\label{eq1}
\end{equation}
where
\begin{equation}
H_L = \sum\limits_{k,\sigma} (\epsilon_{L,k}^0 -ev_L )
a^{\dagger}_{L,k\sigma}a_{L,k\sigma} \ \ ; 
\label{eq2}
\end{equation}
\begin{eqnarray}
H_R &&= \sum\limits_{p,\sigma} \epsilon_{R,p}^0 
a^{\dagger}_{R,p\sigma}a_{R,p\sigma}  
+\sum\limits_p\left[ \Delta^* a_{R,p \downarrow} a_{R,-p \uparrow}   
\right.\nonumber \\
&&\left.+\Delta a^{\dagger}_{R,-p \uparrow} a^{\dagger}_{R,p \downarrow}  
\right] \ \ ;
\end{eqnarray}
\begin{equation} 
H_{d} =\sum\limits_{i,\sigma}\epsilon_i c_{i\sigma}^{\dagger}
c_{i\sigma} ~~;
\end{equation} 
\begin{eqnarray}
H_T &&= \sum\limits_{k,\sigma,i} \left[ t_{L,k i} a^{\dagger}_{L,k\sigma}
c_{i\sigma} + t_{L,k i}^*c_{i\sigma}^{\dagger}a_{L,k\sigma} \right] 
\nonumber \\
&& + \sum\limits_{p,\sigma,i}
\left[ t_{R,k i} e^{iev_R\tau} a^{\dagger}_{R,p\sigma}c_{i\sigma} + 
t_{R,k i}^* e^{-iev_R \tau} c_{i\sigma}^{\dagger}a_{R,p\sigma} \right].
\end{eqnarray}
Here $H_L$ describes noninteracting electrons in the left lead (assumed to 
be a normal-metal), $a_{L,k\sigma}^{\dagger}$ ($a_{L,k\sigma}$) is the 
creation 
(annihilation) operator of electrons in the left lead, and $v_L$ is the 
bias voltage applied on the left lead. $H_R$ describes the right lead which
is a superconducting lead whose gap energy is $\Delta$. $H_d$ is the 
Hamiltonian of the quantum well with multiple discrete energy levels 
characterized by index $i$ and spin $\sigma$. $H_T$ denotes the coupling 
part of the Hamiltonian, $t_{\alpha,k i}$ ($\alpha=L,R$) is the hopping 
matrix and for simplicity is assumed to be independent of the spin index 
$\sigma$. In order to obtain the Hamiltonian (\ref{eq1}) and (\ref{eq2}), 
we have performed a unitary transformation similar to that described in 
Ref.\onlinecite{yeyati1}, so that the bias voltage of the right lead, $v_R$,
appears as a phase factor in the hopping elements. Finally, it is important 
to note that we must include the internal Coulomb potential $U$ inside 
the scattering region for further analysis (see below).

Our analysis follows that of Ref.\onlinecite{sun1} by applying the NEGF 
theory, iterating the equation of motion for the retarded Green's function,
and applying the Keldysh equation for the lesser Green's function at
equilibrium. The current flowing through the normal lead is derived 
to be\cite{sun1} ($e=\hbar=1$),
\begin{equation}
I= I_A + I_1 ~~,
\end{equation}
with  
\begin{eqnarray}
I_A&&=2\int \frac{dE}{2\pi} {\rm Tr} \left[ \Gamma_L G_{12}^r 
\Gamma_L G_{12}^a \right] [ f_L(E+v_L-v_R) \nonumber \\
&& -f_L(E-v_L+v_R)] ~~, 
\label{ia}
\end{eqnarray}
\begin{eqnarray}
I_1 & &=2 \int \frac{dE}{2\pi} \rho_R(E) {\rm Tr} 
\left[ \Gamma_L G_{11}^r \Gamma_R G_{11}^a +
\Gamma_L G_{12}^r \Gamma_R G_{12}^a \right. \nonumber \\
&& \left.-\frac{\Delta}{|E|} (\Gamma_L G_{11}^r
\Gamma_R G_{12}^a + \Gamma_L G_{12}^r
\Gamma_R G_{11}^a ) \right] \nonumber \\
&& \times [ f_L(E+v_L-v_R)-f_R(E)] 
\label{i1}
\end{eqnarray}
where $G_{11}$ and $G_{12}$ are the matrix elements of the $2\times2$ 
Nambu representation.
Here an important departure from previous theory\cite{sun1} is the
explicit inclusion of the Coulomb potential $U({\bf r})$ in the Green's
functions $G_{11}^r$ and $A^r$\cite{wbg2}:
\begin{eqnarray}
G_{11}^r(E)&=&\left[E-H_{d}+U-v_R - {\bf \Sigma}^r_{11}
\right.\nonumber \\
&-& \left.{\bf \Sigma}_{12}^r A^r {\bf \Sigma}_{21}^r \right]^{-1}
\label{g11}
\end{eqnarray}
\begin{equation}
A^r= \left[ E + H^*_{d} +v_R -U - {\bf \Sigma}^r_{22} \right]^{-1}
\ \ \ .
\label{ar}
\end{equation}
Once the electron and hole Green's functions $G_{11}^r$ and $A^r$ were
obtained, $G_{12}^r$ is calculated by
\begin{equation}
G_{12}^r = G_{11}^r {\bf \Sigma}^r_{12} A^r \ \ .
\label{g12}
\end{equation}
The self energy ${\bf \Sigma}^r={\bf \Sigma}^r_L + {\bf \Sigma}^r_R$
is derived\cite{sun1} to be
\begin{equation}
{\bf \Sigma}^r_L= \left( \begin{array}{ll}
         \Sigma^r_L & ~~~ 0 \\
         0 & ~~~ -\Sigma^a_L
         \end{array}
  \right)
\end{equation}
where $\Sigma^r_\alpha \equiv P_\alpha - i \Gamma_\alpha/2$ is the self
energy of the lead $\alpha$ in the normal case. Here $P_\alpha$ is the 
real part and $\Gamma_\alpha$ is the linewidth function. In the wide
bandwidth limit\cite{foot4}, the self energy for the superconducting 
lead is
\begin{equation}
{\bf \Sigma}^r_R= \frac{i\Gamma_R}{2}\left( \begin{array}{ll}
         -\beta_1 & ~~~ \beta_2 \\
          \beta_2 & ~~~ -\beta_1^*
         \end{array}
  \right)
\end{equation}
where $\beta_1=\kappa/\sqrt{E^2-\Delta^2}$, 
$\beta_2=\Delta/\sqrt{E^2-\Delta^2}$,
$\kappa=E$ when $|E|<\Delta$ and $\kappa=|E|$ otherwise so that the 
electron-hole symmetry is preserved. The dimensionless BCS density of states
(DOS) is given by $\rho_R = \theta(|E|-\Delta) \beta_1$. 

We emphasize that the crucial step in developing the gauge invariant
nonlinear DC theory is to include the {\it internal} potential landscape 
$U({\bf r})$ into the Green's functions self-consistently\cite{wbg2}.  In
this work we deal with it at the Hartree level, hence $U({\bf r})$ 
is determined by the self-consistent Poisson equation
\begin{equation}
\nabla^2 U(x) = -4\pi i (G^<_{11}(E,U))_{xx}
\label{poisson}
\end{equation}
where $G^<_{11}$ is the electron lesser Green's function in real space and 
$x$ labels the three dimensional position. From Ref.\onlinecite{sun1}
$G^<_{11}$ is determined by iterating the Keldysh equation at equilibrium,
the result is\cite{sun1}
\begin{eqnarray}
{\bf G}^<_{11} &=& i\int \frac{dE}{2\pi} \left[
G_{11}^r \Gamma_L G_{11}^a f_L(E+v_L-v_R) 
\right. \nonumber \\
&+&\left.G_{12}^r \Gamma_L G_{21}^a f_L(E-v_L+v_R)
\right] \nonumber \\ 
&+&i\int \frac{dE}{2\pi} \rho_R(E) f_R(E)
\left[ G_{11}^r \Gamma_R G_{11}^a 
+G_{12}^r \Gamma_R G_{21}^a \right.\nonumber \\
&-& \left.\frac{\Delta}{|E|} (G_{11}^r \Gamma_R 
G_{21}^a +G_{12}^r \Gamma_R G_{11}^a) \right]\ .
\label{gl}
\end{eqnarray}

Eqs. (\ref{ia}), (\ref{i1}), and (\ref{poisson}) completely determines 
the nonlinear I-V characteristics of the N-S hybrid system: they form the 
basic equations of the gauge invariant nonlinear theory. The 
self-consistent nature of the problem is clear: one must solve the 
quantum scattering problem 
(the Green's functions) in conjunction with the Poisson equation. It is 
easy to prove that the current expression Eqs.(\ref{ia}) and (\ref{i1}) 
are gauge invariant: shifting the bias potential everywhere by a constant 
$v$, $v_{\alpha}\rightarrow v_{\alpha}+v$ so that $U\rightarrow U+v$ 
(since the boundary condition of $U$ is changed by $v$), $I$ from 
Eqs.(\ref{ia}) and (\ref{i1}) remains the same. Eqs. (\ref{ia}), 
(\ref{i1}), and (\ref{poisson}) also form the basis for numerical analysis
of I-V curves for the N-S system. For instance one can compute the 
various Green's functions ${\bf G}$ and the coupling matrix $\Gamma$ using 
tight-binding models\cite{datta2}, and the Poisson equation
can be efficiently solved in real space by powerful numerical 
techniques\cite{wang2}.

In the simplest approximation, the gauge invariant condition can be 
satisfied by putting a gate voltage $V_g$ as was done in 
Ref.\onlinecite{sun1} so that one treats the system as a three probe 
conductor with external voltages $V_L$, $V_R$ and $V_g$ applied at the 
probes. In general, the internal potential is a nonlinear function 
of $V_\alpha$ (see section IV for details), but as a first approximation 
one expands it in terms of $V_\alpha$ in the small voltage limit,
\begin{equation}
U= u_L V_L + u_R V_R + u_g V_g
\end{equation}
where $u_\alpha({\bf r})$ is the characteristic potential which satisfies 
the sum rule $\sum_\alpha u_\alpha=1$. If one makes\cite{sun1} a 
further approximation by assuming $u_L=u_R=0$, the sum rule gives $u_g=1$ 
and $U=V_g$, {\it i.e.} $U$ is just a constant potential shift under these
approximations. 

In distinct contrast to the constant $U$ model, the theory presented
in this section is a microscopic gauge invariant theory. Furthermore, 
in order to discuss charge polarization and electrochemical capacitance 
in the presence of transport, one has to include the self-consistent 
Hartree field rather than just include a constant gate voltage: one can
easily confirm that the constant $U$ model corresponds to the local
charge neutrality approximation, it will therefore not give rise 
to any charge polarization. 

\section{The scattering matrix and Fisher-Lee relation}

The Fisher-Lee relation\cite{fisher-lee} relates scattering matrix of a
conductor to its Green's function. This relation has been widely used 
in analyzing quantum transport through mesoscopic normal 
conductors\cite{datta}. To make a formal connection of our theory presented
in the last section and the scattering matrix theory for N-S systems, 
it is necessary to derive a Fisher-Lee relation for the N-S system.
In this section we provide such a derivation which allows us to make 
comparison between our theory and SMT. Because results (gauge invariant) 
of SMT for N-S system are only known to linear order in bias, we
reduce our theory to that limit in this section. In addition, the
SMT\cite{but3,beenakker} has so far dealt with the sub-gap region 
($|E|<\Delta$) hence we will focus on the Andreev current $I_A$ only, 
although our theory is applicable beyond this limitation as shown by 
Eq. (\ref{i1}).

To prove that our theory reduces to that of the SMT\cite{but3}, we go 
to the wide bandwidth\cite{jauho1} limit at zero bias within the gap. In 
this limit we have\cite{wbg2}
$\Sigma^r_\alpha= -i\Gamma_\alpha/2$, hence Eqs. (\ref{g11}, \ref{ar},
\ref{g12}) become
\begin{equation}
G^r_{11} = \frac{E + H_{d}^* +i\Gamma_L/2+i\Gamma_R \beta_1/2}{X}\ ,
\label{g11a}
\end{equation}
\begin{equation}
G^r_{22} = \frac{E - H_{d} +i\Gamma_L/2+i\Gamma_R \beta_1/2}{X}\ ,
\label{g22a}
\end{equation}
and
\begin{equation}
G^r_{12} = G^r_{21} = \frac{i\Gamma_R \beta_2/2}{X}\ ,
\label{g12a}
\end{equation}
where
\begin{eqnarray}
X&\equiv&(E + H^*_{d} +i\Gamma_L/2+i\Gamma_R \beta_1/2)(E - H_{t}
\nonumber \\
&+&i\Gamma_L/2 +i\Gamma_R \beta_1/2)+ \beta_2^2 \Gamma_R^2/4 \ .
\end{eqnarray}
Note that these reduced Green's functions (\ref{g11a},\ref{g22a},\ref{g12a})
can be collectively defined by
\begin{equation}
{\bf G}^r = \frac{1}{E -H_{eff}+i\Gamma_L/2+i\Gamma_R \beta_1/2}
\label{geff}
\end{equation}
for which the effective Hamiltonian\cite{been2} is given by
\begin{equation}
H_{eff}= \left( \begin{array}{ll}
         H_{d} & ~~~ i\Gamma_R \beta_2/2\\
         i\Gamma_R \beta_2/2 & ~~~ -H^*_{d}
         \end{array}
  \right)\ \ .
\label{heff}
\end{equation}

The scattering matrix approach of Ref.\onlinecite{but3} is recovered by 
letting $E \rightarrow 0$ (which is equivalent to setting the parameter of
scattering theory $\alpha=-i$, see Ref.\onlinecite{but3}), therefore
$\beta_1=0$ and $\beta_2=-i$. We thus obtain
\begin{equation}
H_{eff}= \left( \begin{array}{ll}
         H_{d} ~~~~~ & \Gamma_R/2 \\
         \Gamma_R/2 & -H^*_{d}
         \end{array}
  \right)
\end{equation}
which is exactly the same as that of the scattering matrix theory of 
Ref.\onlinecite{but3}. We thus conclude that the wide bandwidth limit
of our theory is equivalent to SMT at zero bias.

The equivalence of our theory and SMT can be further discussed through the
Fisher-Lee relation. This relation was first derived\cite{fisher-lee} for
normal conductors, and here we derive it for N-S systems within the sub-gap
region. For simplicity we assume that the scattering matrix in the normal 
side of the N-S system is of the Breit-Wigner form:
\begin{equation}
s_{\alpha \beta} = \delta_{\alpha \beta} - i\frac{\sqrt{\Gamma_\alpha 
\Gamma_\beta}}{E - E_0 + i\Gamma/2}
\label{a1}
\end{equation}
with $\Gamma=\Gamma_1+\Gamma_2$. From Ref.\onlinecite{beenakker}, the 
scattering matrix of the N-S hybrid system is given by
\[ 
s_{he}(E) = \alpha e^{-i\phi} s^*_{12}(-E) M_e s_{21}(E)
\]
with
\[
M_e = [1-\alpha^2 s_{22}(E) s^*_{22}(-E)]^{-1}
\]
and 
\[
\alpha=\frac{E - \sqrt{E^2 - \Delta^2}}{\Delta}\ \ .
\]

Using Eq.(\ref{a1}) we obtain 
\[
s_{22}(E) s^*_{22}(-E) = 1 - \frac{-\Gamma_1 \Gamma_2
+ 2i E \Gamma_2}{(E - E_0 +i\Gamma/2)(E+E_0 +i\Gamma/2)}
\]
and
\[
s_{21}(E) s^*_{12}(-E) = \frac{\Gamma_1 \Gamma_2}
{(E - E_0 +i\Gamma/2)(E+E_0 +i\Gamma/2)}\ \ .
\]
Hence 
\[
s_{he} = \alpha e^{-i\phi} \frac{\Gamma_1 \Gamma_2}{Y}
\]
where
\begin{eqnarray}
Y&=&(1-\alpha^2)(E-E_0+i\Gamma/2)(E+E_0+i\Gamma/2) \nonumber \\
&-& \alpha^2 (\Gamma_1 \Gamma_2 - 2i E \Gamma_2) \ \ . \nonumber
\end{eqnarray}

Finally, note that $1-\alpha^2 = 2\alpha \sqrt{E^2 - \Delta^2}/\Delta$, we 
obtain a relationship between the scattering matrix $s_{he}$ and Green's
function $G_{12}^r$,
\begin{eqnarray}
s_{he}(E) & =& \frac{\Gamma_1 \Gamma_2 \Delta}
{2\sqrt{E^2 - \Delta^2}} e^{-i\phi} [(E -E_0 +i\frac{\Gamma_1}{2})
(E+E_0 +i\frac{\Gamma_1}{2}) \nonumber \\
&-& \Gamma_2^2/4 + (2E + i\Gamma_1)
(1+\alpha \frac{\Delta}{\sqrt{E^2-\Delta^2}}) i\frac{\Gamma_2}{2}]^{-1}
\nonumber 
\\
&=& \frac{\Gamma_1 \Gamma_2 \Delta}
{2\sqrt{E^2 - \Delta^2}} e^{-i\phi}/X \nonumber \\
&=& \Gamma_L e^{-i\phi} G^r_{12}
\label{fisher}
\end{eqnarray}
where we have used the fact that $1+\alpha \Delta /\sqrt{E^2 - 
\Delta^2} = E/\sqrt{E^2 - \Delta^2}$, $\Gamma_1=\Gamma_L$,
$\Gamma_2=\Gamma_R$, and Eq. (\ref{g12a}) for $G_{12}^r$.  
Eq.(\ref{fisher}) is the Fisher-Lee relation\cite{fisher-lee} for the
N-S system.
Using Eq. (\ref{fisher}) it is straightforward to show that when the 
Fermi energy is inside the gap, Eq.(\ref{ia}) gives the same result
as that of SMT in Ref.\onlinecite{beenakker,lambert} with $\alpha = 
\exp(-i\arccos E/\Delta)$. 

\section{Applications}

In this section we present detailed analysis for a number of situations 
where analytical expressions can be obtained in closed form. These are
resonance Andreev reflection coefficient at the linear regime; the second
order weakly nonlinear conductance; and the nonequilibrium charge 
distribution.

\subsection{Resonant Andreev reflection}

The phenomenon of resonant Andreev reflection\cite{sun1} is the situation
where the Andreev current is dominated by a resonance transmission through
a level $E_o$ inside the quantum well. For this case the Andreev reflection
coefficient $T_A(E)=Tr[\Gamma_L G_{12}^r\Gamma_L G_{12}^a]$ can be derived
assuming a Breit-Wigner form of the scattering matrix (\ref{a1}). From
Eq.(\ref{fisher}) and note that at sub-gap region energies $E_o$ and $E$
are small therefore we can take $\alpha \approx -i$,  we obtain
\[
G^r_{12} = \frac{i \Gamma_2}{2(E^2-E_0^2)-(\Gamma^2+\delta 
\Gamma^2)/4 + i E (\Gamma+\delta \Gamma)}
\]
where $\delta \Gamma\equiv \Gamma_1-\Gamma_2$. After simple algebra
the Andreev reflection $T_A$ is found to be 
\begin{equation}
T_A = \frac{\Gamma_1^2 \Gamma_2^2}{4(E^2 - E_0^2 + \Gamma\delta 
\Gamma/4)^2 + \Gamma_1^2 \Gamma_2^2 + E_0^2(\Gamma^2 + \delta \Gamma^2)}
\ .
\label{TA}
\end{equation}
This is the Breit-Wigner formula for N-S system. Note that this expression
is different from the formula of Ref.\onlinecite{beenakker} which is
only valid at $E=0$. Several interesting observations of (\ref{TA})
are in order. First, let's consider resonance energy being at $E_o=0$.  
For this situation we have
\[
T_A = \frac{\Gamma_1^2 \Gamma_2^2}{4(E^2 + \Gamma\delta \Gamma/4)^2 + 
\Gamma_1^2 \Gamma_2^2}]\ \ .
\]
Therefore: (i) if $\Gamma_1<\Gamma_2$ or $\delta \Gamma<0$, there are 
two peaks with $T_A=1$ at $E= \pm \sqrt{-\Gamma \delta \Gamma}/2$; 
(ii) when $\Gamma_1=\Gamma_2$, we have  
$T_A = \Gamma_1^2 \Gamma_2^2/[4E^4 + \Gamma_1^2 \Gamma_2^2]$, so that there
is only one peak with $T_A=1$ at $E=0$; (iii). for $\Gamma_1> \Gamma_2$, 
there is only one peak with $T_A = \Gamma_1^2 \Gamma_2^2/[\Gamma^2 \delta 
\Gamma^2/4 + \Gamma_1^2 \Gamma_2^2]<1$ at $E=0$.  Hence when energy level 
of the quantum well is aligned at the center of the superconducting gap, the
resonance Andreev reflection is characterized by one or two peaks depending
on how the normal and superconducting leads are coupled to the quantum well.
Second, when $E_o$ is nonzero, we have:  (i) $\Gamma_1 \le \Gamma_2$, 
two peaks with $T_A = \Gamma_1^2 \Gamma_2^2/[\Gamma_1^2 \Gamma_2^2 + 
E_0^2(\Gamma^2 +\delta \Gamma^2)]$ to occur at $E^2= E_0^2-\Gamma \delta 
\Gamma/4$. 
(ii). $\Gamma_1>\Gamma_2$, just one peak at $E=0$ if 
$\Gamma \delta \Gamma/4>E_0^2$, otherwise two peaks at 
$E^2= E_0^2-\Gamma \delta \Gamma/4$. Note that 
when $E_o$ is nonzero, the peak values are always less than one. 

\subsection{Weakly nonlinear regime}

For weak nonlinearity we can expand all quantities in terms of the 
external bias voltage\cite{but1} and obtain results order by order. 
Such an expansion makes sense when bias is finite but small. This
approach was adapted in SMT\cite{but1} and response theory\cite{ma} 
for analyzing normal mesoscopic conductors. For the N-S system we will 
derive formula for the local density of states (LDOS) and the second order
weakly nonlinear DC conductance. These are the interesting quantities 
for weakly nonlinear regime.

In both SMT\cite{but1} and response theory\cite{ma}, LDOS plays a very 
important role. From our NEGF theory LDOS can be easily derived from the
right hand side of Eq. (\ref{poisson}), which is the charge density,
with the help of Eq. (\ref{gl}). Here we shall present the explicit 
expression at the lowest order\cite{ma} expansion in external bias.
Hence we seek the solution of $U({\bf r})$ in the following form,
\begin{equation}
U= \sum_{\alpha} u_{\alpha} v_{\alpha} +\frac{1}{2}\sum_{\alpha 
\beta} u_{\alpha \beta} v_{\alpha} v_{\beta} + ...
\label{char}
\end{equation}
where $u_\alpha({\bf r})$ and 
$u_{\alpha \beta ..}({\bf r})$ are the characteristic 
potentials\cite{but1,ma}. It can be shown that the characteristic
potential satisfies many sum rules\cite{but1,ma},
$\sum_{\alpha} u_{\alpha} =1$
and 
$\sum_{\gamma\in\beta} u_{\alpha\{\beta\}_l}=0$,
where the subscript $\{\beta\}_l$ is a short notation of $l$ indices
$\gamma,\delta,\eta,\cdot\cdot\cdot$. Expanding $G^<_{11}$ of Eq. 
(\ref{poisson}) in power series of $v_\alpha$\cite{foot}, we can derive 
equations for all the characteristic potentials. In particular the 
expansions are facilitated by iterating the following Dyson equation 
to the appropriate order 
\[
A^r=A^r_0 - A^r ~ (v_R - U) ~ A^r_0 
\]
and 
\begin{eqnarray}
G^r_{11} &=& G^r_{11,0} - G^r_{11} ~ (U-v_R) ~ G^r_{11,0} + 
G^r_{11} ~ {\bf \Sigma}^r_{12} ~ (A^r-A^r_0) \nonumber \\
& & ~ {\bf \Sigma}^r_{21} ~ G^r_{11,0}
\label{expand}
\end{eqnarray}
where $A^r_0$ and $G^r_{11,0}$ are equilibrium hole and electron
Green's functions. The expansion of $G^r_{12}$ can be made similarly. 
At the lowest order, we thus obtain the local charge density in the 
presence of transport\cite{foot1},
\begin{equation}
\rho(x) = i (G^<_{11} - G^<_{11,0})_{xx} = \rho_{inj}+\rho_{ind}
\label{r1}
\end{equation}
where 
\begin{eqnarray}
\rho_{inj}&=& (dn_e/dE-dn_h/dE)(v_L-v_R)\nonumber \\
&-& (1/2) (d^2n_e/dE^2+d^2n_h/dE^2)(v_L-v_R)^2
\end{eqnarray}
is the injected charge from the normal lead. $dn_e/dE$ is the injectivity of 
electron, {\it i.e.} the DOS for an electron coming from left lead and 
exiting the system as an electron, 
\begin{equation}
dn_e(x)/dE=\int (dE/2\pi) (-\partial_E f_L) (G^r_{11,0} 
\Gamma_L G^a_{11,0})_{xx} \ .
\label{dnee}
\end{equation}
In addition $dn_h/dE$ is the injectivity of a hole, {\it i.e.} 
the DOS for a hole coming from left lead and exiting the system as an 
electron,
\begin{equation}
dn_h(x)/dE=\int (dE/2\pi) (-\partial_E f_L) (G^r_{12,0} 
\Gamma_L G^a_{12,0})_{xx} \ .
\label{dneh}
\end{equation}
Finally $d^2n/dE^2$ is the derivative of $dn/dE$ with respect to energy. 
Note that Eqs.(\ref{dnee}) and (\ref{dneh}) are the same as that defined
in the scattering approach of Gramespacher and B\"uttiker\cite{but3}. 

In Eq.(\ref{r1}) the induced charge due to long range Coulomb interactions
is derived to be given by
\begin{eqnarray}
\rho_{ind}(x) &=& -\int (dE/\pi) f_L {\rm Im} \left[G^r_{11,0} 
(u_L-{\bf \Sigma}^r_{12} A^r_0 u_L A^r_0 {\bf \Sigma}^r_{21} ) \right. 
\nonumber \\
& & \left.G^r_{11,0} \right]_{xx} ~ (v_L-v_R) \nonumber \\
&\equiv&-\sum_{x'} \Pi_{xx'} u_L(x') ~ (v_L-v_R) 
\label{pi}
\end{eqnarray}
where $\Pi$ is the generalized Lindhard function which reduces to the
Lindhard function of normal conductor\cite{levinson,but1,ma} in the 
limit $\Delta \rightarrow 0$. For example, using the wide bandwidth limit 
expressions (such as Eq.(\ref{g11a}) for the Green's function, or using
the Breit-Wigner form Eq.(\ref{a1}) for the scattering matrix,
at small Fermi energy (so that $\beta_1=0$ and $\beta_2=-i$) we can
calculate the Lindhard function $\Pi$ exactly at zero temperature from
its definition above:
\begin{equation}
\Pi = \frac{2}{\Gamma_L \Gamma_R} \left[ \frac{\pi}{2}
- \arctan\frac{2E^2-(\Gamma_L^2-\Gamma_R^2)/2}{\Gamma_L \Gamma_R} \right]
\end{equation}
where we have set the quantum well level $E_o=0$. Hence the Lindhard
function is a smooth function increasing with energy monotonically. 

With these quantities the Poisson equation becomes,
\begin{equation}
-\nabla^2 u_L(x)+ 4\pi \sum_{x'} \Pi_{xx'} u_L(x')
= 4\pi (\frac{dn_e(x)}{dE}-\frac{dn_h(x)}{dE})
\label{poisson1}
\end{equation}
\begin{equation}
-\nabla^2 u_{LL}(x)+ 4\pi \sum_{x'} \Pi_{xx'} u_{LL}(x')
= 4\pi (\frac{d\tilde{n}_e(x)}{dE}-\frac{d\tilde{n}_h(x)}{dE})
\end{equation}
where $d\tilde{n}_e/dE$ and $d\tilde{n}_h/dE$ are the second order
injectivities\cite{ma,zhao}. These partial differential equations can
at least be solved numerically. However to
avoid numerics one may apply the quasi-neutrality approximation\cite{but0} 
by neglecting the spatial derivative in Eq.(\ref{poisson1}). 
This way the characteristic potential is obtained as\cite{foot3}
\begin{equation}
u_L = (\frac{dn_e}{dE}-\frac{dn_h}{dE})/\Pi\ \ .
\label{ul}
\end{equation}

In terms of the characteristic potential we now derive the second order 
nonlinear conductance due to Andreev reflection. In the weakly nonlinear 
regime, only the Andreev current $I_A$ is relevant which can be expanded in 
terms of external bias voltage difference $v\equiv v_L - v_R$,
\[
I_A = G_{11} v + G_{111} v^2 + ...
\]
From this definition of conductance coefficients $G_{11}$ and $G_{111}$, we
expand Eq.(\ref{ia}) in terms of $v$ to obtain, 
\[
G_{11} = 4\int (dE/2\pi) (-\partial_E f_L) T_A
\]
and 
\begin{equation}
G_{111} = -4\int \frac{dE}{2\pi} (-\partial_E f_L) 
Tr\left[\frac{dG_A}{dU} u_L \right]
\label{g111}
\end{equation}
where $G_A \equiv \Gamma_L G^r_{12} \Gamma_L G^a_{12}$ and $dG_A/dU$
is easily calculable using Eq.(\ref{expand}) and the relation in
Ref.\onlinecite{foot1}. To compare with the second order conductance 
of normal conductor $G^N_{111}$, we note that $G^N_{111}$ has two 
contributions\cite{but0,wbg2}. One of them comes from Coulomb interaction,
\begin{eqnarray}
&~& G^N_{111} = \int (dE/2\pi) Tr[ G^a_0 (\Gamma_L G^r_0 u_L +
u_L G^a_0 \Gamma_L \nonumber \\ 
&-& 1/2 \Gamma_L G^r_0 
- 1/2 G^a_0 \Gamma_L) G^r_0 \Gamma_R] \partial_E f 
\label{g111n}
\end{eqnarray}
However, for NS system, if the Coulomb interaction is not important 
(when $u_L=0$), we would have $G_{111}=0$. For example, for an ideal
ballistic wire, or for a symmetric quantum well at a resonant tunneling point,
every incident charge is perfectly Andreev reflected. Therefore, for these
examples no charge accumulation is possible. From Eqs.(\ref{dnee}) and 
(\ref{dneh}), we can easily verify that $dn_e/dE=dn_h/dE$ near a 
resonant point for a symmetric system and hence a vanishing $G_{111}$ since 
$u_L=0$ from Eq.(\ref{ul}). In contrast, when $u_L=0$, $G^N_{111}$
is nonzero from Eq.(\ref{g111n}). 

\subsection{Electrochemical capacitance}

Using the NEGF theory one can also investigate the nonequilibrium charge
distribution inside the N-S system. For this purpose we divide the system 
into two regions: in region I the charge is positive and in region II it 
is negative. The total charge in region I can be calculated using 
Eq.(\ref{r1}): $Q_I = \int_I \rho(x)dx$. Expanding $Q_I$ in powers 
of $v$ in the following form\cite{ma,zhao}
\[
Q_I = C_{11}v+\frac{1}{2} C_{111} v^2 + ... \equiv C(v) v
\]
this defines the electrochemical capacitance coefficients $C_{11}$,
$C_{111}$ and the general voltage dependent electrochemical 
capacitance\cite{apl} $C(v)$. It is not difficult to confirm that the
first two coefficients are 
\begin{eqnarray}
C_{11} &=& \int_I dx (\frac{dn_e}{dE}-\frac{dn_h}{dE}) \nonumber \\
&-& \int_I dx dx' \Pi(x,x') u_L(x')
\label{c11}
\end{eqnarray}
\begin{eqnarray}
C_{111} &=& \int_I dx (\frac{d\tilde{n}_e}{dE}-\frac{d\tilde{n}_h}{dE}) 
\nonumber \\
&-& \int_I dx dx' \Pi(x,x') u_{LL}(x')\ \ .
\label{c111}
\end{eqnarray}

To get some physical insight for these coefficients we consider the 
discrete potential model\cite{but4}. In addition, we parameterize the
characteristic potentials by the geometric capacitance $C_0$, in terms of 
which the Poisson equation becomes
\begin{equation}
C_0 (u_I - u_{II}) = (D^e_I-D^h_I) v - \Pi_I u_I = C_{11} v
\label{x1}
\end{equation}
\begin{equation}
-C_0 (u_I - u_{II}) = (D^e_{II}-D^h_{II}) v - \Pi_{II} u_{II}
\label{x2}
\end{equation}
where we have set $D^e=\int_I dx(dn_e/dE)$, $D^h=\int_I dx(dn_h/dE)$, 
$\Pi_I = \int_I dx \Pi(x,x)$ and small bias limit is assumed. We solve 
the characteristic potentials $u_L$ and $u_{LL}$ through these two 
equations in terms of $C_0$. This leads to the following expression for the
electrochemical capacitance coefficient $C_{11}$ for a N-S system:
\begin{equation}
C_{11} = \frac{(D^e_I-D^h_I)/\Pi_I - (D^e_{II} - D^h_{II})/\Pi_{II}}
{C_0^{-1} + \Pi_I^{-1} + \Pi_{II}^{-1}}\ \ .
\label{ana}
\end{equation}
In particular, in the limit of gap $\Delta \rightarrow 0$, from 
Eqs.(\ref{dnee}), (\ref{dneh}), and (\ref{pi}) we obtain $D^h=0$, 
$\Pi=dn/dE$, and $D^e=dn_L/dE$ where $dn_L/dE$ is the injectivity of 
left lead. In this situation Eq.(\ref{ana}) reduces to the expression of the 
electrochemical capacitance for a normal conductor\cite{zhao}.
Let's consider a symmetric tunneling N-S system.
At the resonant point, the electron will be reflected as the hole due 
to the Andreev reflection. As a result, the capacitance $C_{11}$ 
vanishes since $D^e=D^h$ and there is no charge accumulation. 

\section{Summary}

In this work we have developed a gauge invariant NEGF theory for hybrid 
N-S systems.  This theory explicitly takes into account the long range
Coulomb interaction in the normal region. When the Fermi energy is
inside the gap, we have shown that at linear regime this theory is equivalent
to the scattering matrix theory. We have also derived the Fisher-Lee
relation for the N-S system hence a formal connection between our NEGF theory
and the SMT is made for these systems. Because of gauge invariance, our
theory is applicable for nonlinear regime for which we have derived
an explicit expression for nonlinear current-voltage characteristics for N-S
devices. This result can be further simplified in the weakly 
nonlinear regime, for which we have analyzed the second order nonlinear 
conductance and the generalized Lindhard function. It is interesting the see
that for N-S systems the concept of injectivity is naturally extended to
include the injectivity of holes: these quantities automatically appear 
in our formalism. Our theory included charge polarization effect hence 
can be applied to analyze the electrochemical capacitance coefficients at 
the linear and nonlinear orders in bias. In particular we have derived 
an analytic expression of the linear electrochemical capacitance of N-S 
system within the discrete potential model. 

While this paper concentrated on the development of a theoretical formalism
in terms of the Green's functions, it is obvious that numerical
computations can be carried out applying the analytical expressions derived
here. This way one can avoid the various approximations used here in order
to obtain closed form results. Of particular interests are the investigation
of nonlinear I-V curves without the wide bandwidth approximation; the 
calculation of nonlinear conductance coefficients without the 
quasi-neutrality approximation; and the study of nonequilibrium charge 
distribution without the discrete potential approximation. These, however, 
will be the subject of a future report.

\bigskip 
{\bf Acknowledgments.}
We gratefully acknowledge support by a RGC grant from the SAR Government 
of Hong Kong under grant number HKU 7215/99P and a CRCG grant from the 
University of Hong Kong. H.G. is supported by NSERC of Canada and FCAR 
of Qu\'ebec. T.H.L is supported by the research grant from the Chinese 
National Natural Science Foundation and the State Key Laboratory for
Mesoscopic Physics in Peking University.


\begin{thebibliography}{00}
\bibitem{foot}
Note that only the first two terms are needed in expanding Eq.(\ref{gl}) 
at small vias voltages.
\bibitem{foot1}
Here we have applied the sum rules of the characteristic potential  
and used the fact that
$G^r_{11} \Gamma_L G^a_{11} +G^r_{12} \Gamma_L G^a_{12} =
i G^r_{11}-i G^a_{11} \equiv -2{\rm Im} G^r_{11}$.
\bibitem{foot3}
Here we have used the Thomas-Fermi approximation. See
Ref.\onlinecite{butz} and Ref.\onlinecite{wbg2} for discussions of
this approximation.
\end{thebibliography}
\end{document}